\documentclass[12pt]{article}
\usepackage{pdproc,epsfig} 
\newcommand{\quq}{\theta_{13}}
\newcommand{\nmt}{\nu_\mu \rightarrow \nu_\tau}
\newcommand{\nme}{\nu_\mu \rightarrow \nu_e}

\newcommand{\dmsq}{\Delta m^2}

\begin{document}

\renewcommand{\thefootnote}{\alph{footnote}}
  
\title{NEW PROJECTS IN UNDERGROUND PHYSICS
}
\author{ MAURY GOODMAN}
\address{ High Energy Physics Division
\\
Argonne National Laboratory
\\
Argonne IL 60439\\
 {\rm E-mail: maury.goodman@anl.gov}}
\abstract{A large fraction of neutrino research is taking place in
facilities underground.  In this paper, I review the underground facilities
for neutrino research.  Then I discuss ideas for future reactor experiments being 
considered
to measure $\quq$ and the UNO
proton decay project.}
\normalsize\baselineskip=15pt

\section{Introduction}
Large numbers of particle physicists first went underground in the early
1980's to search for nucleon decay.  Atmospheric neutrinos
were a background to those experiments, but the study of atmospheric
neutrinos has spearheaded tremendous progress in our understanding
of the neutrino.  Since neutrino cross sections, and hence event rates are
fairly small, and backgrounds from cosmic rays often need to be minimized
to measure a signal, many more other neutrino experiments are found
underground.  This includes experiments to measure solar $\nu$'s,
atmospheric $\nu$'s, reactor~$\nu$'s, accelerator $\nu$'s and
neutrinoless double beta decay.
\par

In the last few years we have seen remarkable progress in understanding 
the neutrino.
Compelling evidence for the existence of neutrino mixing and 
oscillations has been 
presented by Super-Kamiokande\cite{bib:sk1} in 1998, based on
 the flavor ratio and
zenith angle distribution of atmospheric neutrinos.  That 
interpretation is supported
by analyses of similar data\cite{bib:sim} from IMB, 
Kamiokande, Soudan 2 and MACRO.
And 2002 was a miracle year for neutrinos, with the results 
from SNO\cite{bib:sno} and KamLAND\cite{bib:kamland} solving the long
standing solar neutrino puzzle and providing evidence for 
neutrino oscillations using
both neutrinos from the sun and neutrinos from 
nuclear reactors.  Finally, the K2K long-baseline
neutrino oscillation experiment\cite{bib:k2k} has the first
indications that accelerator
neutrinos also oscillate.
\par In preparation for this meeting, I asked the previous speaker,
Sandip Pakvasa from the University of Hawaii, ``What are the
outstanding issues in the neutrino sector for future experiments
to address?"  His list was:
\begin{enumerate}
\item 
See the dip in atmospheric $\nu$ L/E distribution
\item 
Measure the whole solar $\nu$ energy spectrum.
\item 
Determine the value of Ue3 ($\quq$).
\item 
Are CP violation effects large?
\item 
If so, what is the CP phase $\delta$?
\item 
Can we see the $\tau$'s in $\nmt$ oscillation?
\item 
Can we settle the LSND question?
\item 
What are  the absolute $\nu$ masses?
\item 
Is the neutrino Majorana or Dirac?
\item 
Is the mass hierarchy normal or inverse?
\item 
Are there astrophysical sources of $>$TeV $\nu$'s?
\end{enumerate}
\par Indeed, those questions are just the ones that have motivated
neutrino physicists, and there are a plethora of new proposals for
future projects at underground laboratories and accelerators.  These
include the study of solar, atmospheric, reactor, accelerator and
astrophysical sources of neutrinos.  There is also the direct search
for neutrino mass in Tritium beta decay, and the search for
Majorana masses in neutrinoless double beta decay.
\par In this paper I will review some underground
projects focusing on those not found
elsewhere on the agenda.  I will start by discussing
 major existing and proposed 
facilities where underground physics research takes place.
In Sections 3 and 4 I will briefly comment on plans for new real time
 solar neutrino experiments and site issues for future off-axis
long-baseline neutrino experiments to measure $\quq$.\cite{bib:harris}
In Section 5 I will present my thoughts about new reactor neutrino
experiments to measure $\quq$.  In Section 6 I will return to
nucleon decay and describe the UNO experiment.

\section{Underground Laboratories for Neutrino Research}

\par The nicest facility for underground physics is located at the
Gran Sasso Laboratory in a mountain road tunnel in Italy.  It is
operated there by INFN for experimentalists from throughout the
world.  There are three large halls in which a number of neutrino
experiments have operated and are being built.  These include the
Gallex and GNO experiments which have
measured the pp solar neutrinos; the MACRO experiment which measured
the angular distribution of atmospheric neutrinos; Borexino which
is being built to measure the solar neutrinos from $^7Be$; the
Heidelberg-Moscow neutrino-less $\beta\beta$ decay experiment on
Germanium, which is a the most sensitive search to date; the LVD
detector which is searching for neutrinos from supernovae; and the
OPERA and ICARUS projects which will measure neutrinos associated
with the CERN Neutrinos to the Gran Sasso (CNGS) program starting
in 2006.

\par The Kamioka mine is located in the Japanese Alps on the western
side of Japan.  It is home to the
50 kiloton Super-Kamiokande water Cerenkov detector.  In
November 2002 it started running again, with half the photo-tube
coverage that it had before its accident in December 2001.  Nearby
is the KamLAND detector which looks at reactor neutrino
disappearance from 26 reactors throughout Japan.  There are also
early plans to put the one megaton Hyper-Kamiokande experiment
in a cavern in the Tochibora mine, about 3 km south of the present
Mozumi mine.  That could pair with the present Super-Kamiokande
facility as an off-axis site from the new JPARC accelerator at
Tokai.
\par The Baksan facility is located in the Caucasus mountains
in southern Russia.  The Gallium solar $\nu$ experiment SAGE
is located in a deep section with a minimum  overburden 4700 MWE.  The
Gallium experiment was crucial in confirming the solar neutrino
deficit.  Also
located at Baksan is the 4 layer Baksan Underground Scintillation
Telescope (BUST) located at an minimum overburden of 850 MWE.  BUST has
been studying the zenith angle distribution of upward atmospheric
induced neutrino events, which is relevant to the atmospheric
neutrino deficit.

\par The Sudbury Neutrino Observatory (SNO) is taking 
data that has provided revolutionary insight 
into the properties of neutrinos and the core of 
the sun. 
The detector is in 
INCO's Creighton mine near Sudbury, Ontario. SNO 
uses 1000 tons 
of heavy water, on loan from Atomic Energy of 
Canada Limited (AECL), contained in a 12 meter 
diameter acrylic vessel. Neutrinos which react with
     the heavy water (D2O) 
are detected by an array of 9600 photomultiplier 
tubes.
The detector laboratory is extremely 
clean to reduce background signals from radioactive elements.
Besides the heavy water detector, the Canadian government has recently
funded a new international facility for underground science called
SNOLAB which will provide a low background facility nearby.

\par The Soudan Underground Physics Laboratory is located in a facility
maintained by the State of Minnesota Department of Natural Resources
and operated as a tourist attraction as part of northern Minnesota's
iron range.  Two connected laboratory spaces are located on the 27th
level with an overburden of 2100 MWE.  The nucleon decay experiment
Soudan 2\cite{bib:s2} operated in one hall from 1986-2001.  Also in 
that hall the Cryogenic Dark Matter Search (CDMS) is being installed.
The 5.6 kton MINOS iron toroid detector which will serve as the far
detector for the Fermilab long-baseline neutrino experiment is 96\%
installed as of May 2003.

\par The Frejus laboratory is in a tunnel between France and Italy.
It was also the site of a proton decay experiment which ran from 1984-1990.
The construction of a parallel safety tunnel has been motivated by the
2000 accident in the Mont Blanc tunnel.  That provides the opportunity
for considering construction of a large new scientific laboratory space
that could be used for a large new detector such as UNO.  Such a detector,
190 km from CERN, would be useful in conjunction with a new neutrino
superbeam from CERN using the proposed Superconducting Proton LINAC (SPL)
or a beta beam using the SPL injecting $^6$He or $^{18}$Ne from
ISOLDE into the SPS and a new storage ring.\cite{bib:beta}

\par 
The Waste Isolation Pilot Plant, or WIPP facility, in 
Carlsbad New Mexico was built to 
store low-level radioactive waste from the US military.
It is the world's first
 underground repository licensed to safely and permanently
 dispose of transuranic radioactive waste left from the research
 and production of nuclear
 weapons.  After more than
 20 years of scientific study,
 public input, and regulatory
 struggles, WIPP began
 operations on March 26,
 1999.   It is considered as a possible site for the supernova experiment
OMNIS which will feature a measurement of the neutral current events
from supernovae; the 400 kiloton proton decay experiment UNO; and
EXO, a new neutrino-less $\beta\beta$ decay experiment which uses Xenon.
Experimental facilities located in the salt are sufficiently removed
from the radioactive waste that they present no backgrounds to any of
these experiments.

\par There is considerable interest in the United States in developing
a National Underground Laboratory Facility (NUSEL) for future neutrino 
experiments.  A panel was appointed in 2002 by the U.S. National
Research Council to study future neutrino facilities, with an 
emphasis on NUSEL and also a neutrino telescope ICE-CUBE.  In its
conclusion section, they wrote\cite{bib:nufac}:
``In summary, our assessment is that a deep underground laboratory in 
the US. can house a
new generation of experiments that will advance our 
understanding of the fundamental properties
of neutrinos and the forces that govern the 
elementary particles, as well as shedding light on the
nature of the dark matter that holds the Universe together. 
Recent discoveries about neutrinos, as
well as new ideas and technologies make possible 
a broad and rich experimental program.
Considering the commitment of the U.S. community 
and the existing scientific leadership in this
field, the time is ripe to build such a unique facility."
The favored location for NUSEL is the Homestake South Dakota 
mine where the famous Davis experiment ran for many years.
There is also a proposal to locate the laboratory with
horizontal access in a new
facility near San Jacinto California.

\par India was home to one of the earliest and deepest underground
facilities with the KGF mine and experiment from the 60's through
the early 90's.  Now a group from several institutes throughout
India is studying the possibility of a new neutrino observatory
to measure atmospheric neutrinos and be the possible site for a
long-baseline neutrino program from a neutrino factory.  
Two sites which have hydro-electric projects near large hills are
being considered.  The PUSHEP site (Pykara Ultimate Stage Hydro
Electric Project) near Ooty in Tamil Nadu is located in a vein of
high quality rock, and is also close to a high altitude cosmic
ray facility.  The RAMMAM Hydro Electric Project site is located
in the Himalayas near Darjeeling, and has the possibility of a
laboratory with a much greater overburden, suitable for solar neutrino
experiments.  A project team has gotten a grant to study both
sites and design a 30 kiloton Iron/RPC calorimeter for atmospheric
neutrinos.
\section{Real Time Solar Neutrino Experiments}
Ray Davis, a winner of the 2002 Nobel prize in physics, started
underground physics in the 1960's with his Homestake solar neutrino
experiment.  Solar neutrino experiments have the most stringent
background requirements to date which has generally made those 
experiments the deepest, and driven the depth requirements for
consideration of a multi-purpose underground laboratory.  
\par Previous solar neutrino experiments have involved chemical
extraction of chlorine or gallium.  The next solar neutrino
experiments will be Borexino and KamLAND,\cite{bib:suzuki} which
will measure solar neutrino experiments in real time.  
\par Ideas for future real-time solar neutrino experiments are listed
in Table \ref{tab:solar}.

\begin{table}[h]
%\hskip4pc\vbox{\columnwidth=26pc
\begin{tabular}{|l|l|l|} \hline
Experiment\cite{bib:noi} & Status & Feature \\  \hline
Borexino &  under construction & scintillator  \\ \hline
KamLAND &  running &  scintillator \\ \hline
HELLAZ & R\&D & Liquid Helium \\ \hline
XMASS &  R\&D &  Liquid Xenon \\ \hline
CLEAN &  R\&D &  Liquid Neon \\ \hline
HERON &  R\&D &  cryogenic \\ \hline
LENS &  R\&D &  $^{176}$Yb \\ \hline
MOON &  R\&D &  $^{100}$Mo \\ \hline
GENIUS &  R\&D &  $\nu e$ in Germanium \\ \hline
\end{tabular}
\caption{Possible Future Real Time Solar Neutrino Experiments}
\label{tab:solar}
\end{table}

\section{Can Off-axis Experiments be at the Surface?}
\par The long-baseline programs that are or soon will be running 
are in Japan (K2K), Europe (CNGS) and the United States (NuMI/MINOS).
The far detectors for all three projects
are located deep enough underground
that there are negligible cosmic rays mimicking beam induced neutrino 
interactions.   There was a previous proposal at Brookhaven\cite{bib:889}
for a long-baseline neutrino experiment on the surface using a water 
Cerenkov Detector.  In their proposal, they performed the only
study of long-baseline neutrino backgrounds in the relevant
energy region of which I'm aware.
\par A new long-baseline experiment is being proposed that
would use a 50 kton detector to search for $\nme$ in the NuMI 
beam\cite{bib:p929} as
evidence for a non-zero value of $\quq$.  The beam spill with one-turn
extraction from the Fermilab Main Injector will be $10\mu s$ long.
Each year, assuming $2\times 10^7$ pulses and an area of about 1000 m$^2$,
there will be 1.2 $\times~10^7$ muons and 8 $\times~10^5$ neutrons 
($E~>~0.1$ GeV) going through
a detector on the surface.  There are four possible handles to reduce
these backgrounds:
\begin{enumerate}
\item overburden
\item active veto
\item pattern recognition
\item bunch timing
\end{enumerate}
The last option would drive up the cost of electronics and cannot reduce
the time window substantially.
 Of course most muons do not look like
contained events, but the existence of cracks or other dead regions in
the detector will make muons a problem.  BNL889 had qualitatively different
pattern recognition issues than the FNAL off-axis experiment will experience, 
but it was going to be 
a homogeneous detector without cracks.  I scale and compare the backgrounds
for illustrative purposes only.  NuMI off-axis will need a rejection of
2.1 $\times~10^6$ for $\mu$'s and $1.6~\times~10^5$  for neutrons, to obtain Signal/Background $>$ 1
for $\sin^2 2\quq = 0.01$.  BNL P889 calculated that they would
achieve a rejection of 7 $\times~10^5$ for $\mu$'s and 3.4 $\times~10^2$
 for neutrons using an active veto
and pattern recognition for electrons 0.3 to 4 GeV.  I conclude that operating
NuMI-off-axis on the surface without overburden will be challenging.

\section{Reactor Experiments}
\par
\ From the discovery of the neutrinos by Reines and Cowan\cite{bib:reines}
at Savannah River to the evidence for $\bar{\nu}_e$ disappearance at
KamLAND\cite{bib:kamland}, reactor neutrino experiments have studied
neutrinos in the same way -- observation of inverse beta decay with
scintillator detectors.  Since the signal from a reactor falls with
distance L as 1/L$^2$, as detectors have moved further away from
the reactors over the years, it has become more important to reduce
backgrounds.  That can be done by putting experiments underground;
and experiments one kilometer or more away from reactors (Chooz, Palo Verde
and KamLAND) have been
underground.
\par The KamLAND experiment measured a 40\% disappearance of
$\bar{\nu}_e$ presumably associated with the 2nd term in Equation \ref{eq:p}:
\begin{equation}
\label{eq:p}
P(\bar{\nu}_e \rightarrow \bar{\nu}_e) \cong 
- \sin^2 2 \theta_{13} \sin^2(\Delta m^2_{atm} L/4E)  
- \cos^4 \quq \sin^2 2 \theta_{12} \sin^2(\Delta
m^2_{12} L/4E) + 1
\end{equation}
The Chooz and Palo Verde data put a limit on $\quq$ (through the first
term in Equation \ref{eq:p}) of $\sin^2 2 \theta_{13} < 0.1$.  
Those experiments could not have had greatly improved sensitivity
to $\quq$ because of uncertainties related to knowledge of the flux of 
neutrinos from the reactors.  They were designed to test
whether the atmospheric neutrino anomaly might have been due to
$\nme$ oscillations, and hence were searching for large mixing.
\par Any new experiment to look for non-zero values of $\quq$ would
need the following properties:
\begin{itemize}
\item two or more detectors to reduce uncertainties to the reactor flux
\item identical detectors to reduce systematic errors related to detector
acceptance
\item carefully controlled energy calibration
\item low backgrounds and/or reactor-off data
\end{itemize}
\par
\begin{figure}
\vspace*{13pt}
%\leftline{\hfill\vbox{\hrule width 5cm height0.001pt}\hfill}
\begin{center}
         \mbox{\epsfig{figure=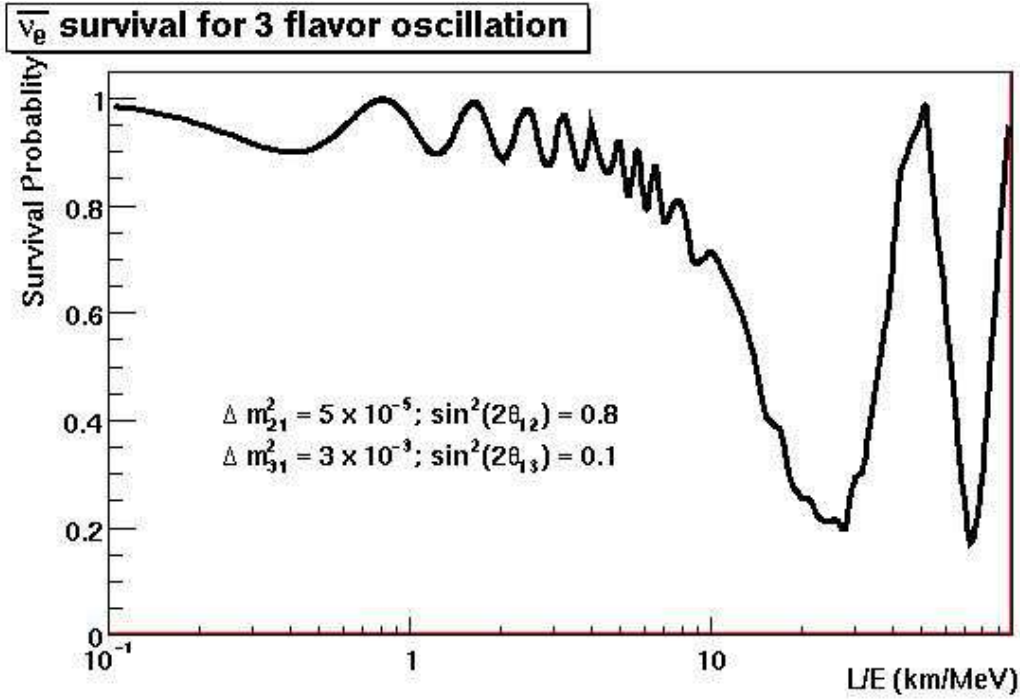,width=15.0cm}}
%\leftline{\hfill\vbox{\hrule width 5cm height0.001pt}\hfill}
\caption{Probability of $\nu_e$ disappearance versus L/E for 
$\quq$ at its current upper limit}
\label{fig:P}
\end{center}
\end{figure}

 In Equation \ref{eq:p}, the values of $\theta_{12}$, $\Delta m^2_{12}$
and $\Delta m^2_{atmo}$ are approximately known.  In Figure \ref{fig:P},
The probability of $\bar{\nu}_e$ disappearance as a function of L/E
is plotted with  $\quq$ 
assumed to be at its maximum allowed value.  Note that
CP violation does not affect a disappearance experiment, and that
matter effects can be safely ignored in a reactor experiment.  The
large variation in P for L/E$>$ 10 km/MeV is the effect seen by
KamLAND and solar $\nu$ experiments.  The much smaller deviations from
unity for L/E $<$ 1 km/MeV are the goal for an accurate new reactor
experiment.

\par
\begin{figure}
\vspace*{13pt}
\leftline{\hfill\vbox{\hrule width 5cm height0.001pt}\hfill}
         \mbox{\epsfig{figure=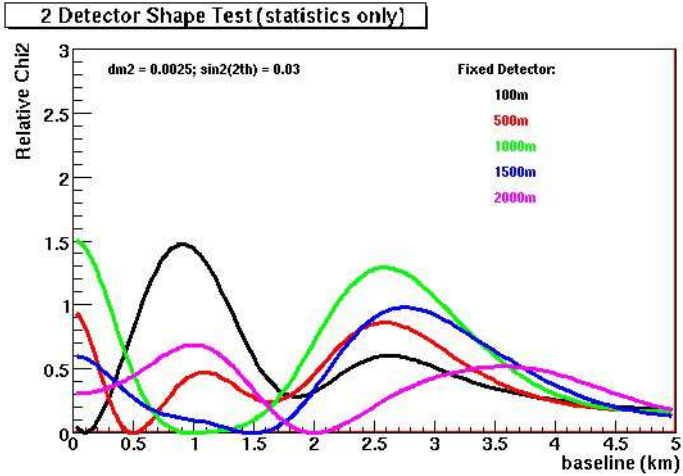,width=10.0cm}}
\leftline{\hfill\vbox{\hrule width 5cm height0.001pt}\hfill}
\caption{$\Delta \chi^2$ for two detectors versus detector location for
statistics only.  Other oscillation parameters have been fixed.
Locations at 100 m and 1000 m are preferred.}
\label{fig:r1}
\end{figure}

\begin{figure}
\leftline{\hfill\vbox{\hrule width 5cm height0.001pt}\hfill}
         \mbox{\epsfig{figure=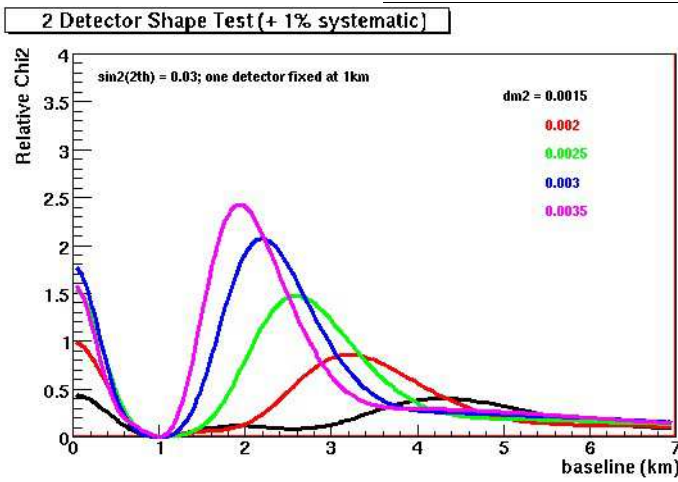,width=10.0cm}}
\leftline{\hfill\vbox{\hrule width 5cm height0.001pt}\hfill}
\caption{ $\Delta \chi^2$ for two detectors with one detector fixed at 1000m
and the other location varied.  A 1\% bin-to-bin systematic error is included,
but the $\Delta m^2_{atm}$ has been allowed to vary from 1.5 to 3.5 
$\times~10^{-3}~eV^2$}
\label{fig:r2}
\end{figure}

 The optimization of detector distances for such a new experiment
is straightforward.  The statistical power comes from measuring a
deficit of $\bar{\nu}_e$ (up to a few percent) at the far detector,
along with a change in the energy spectrum consistent with that deficit.
Up to systematic errors in the rate and energy spectrum, one can construct
a $\chi^2$ difference between a near detector and a far detector.  
In Figure \ref{fig:r1}, the detector location has been fixed for each
curve and the statistical power of these tests calculated as the location
of the second detector is varied.  In Figure \ref{fig:r2}, one
detector has been fixed at 1000 m while the location of the second detector
is varied with a
1\% systematic error folded in.
  The different curves in that plot reflect uncertainty in the
parameter $\Delta m^2_{atm}$.   The optimum locations for detectors is thus
sensitive to eventual systematic errors as well as oscillation parameters.
But a near detector around 100m and a far detector around 1000 m will be
near the optimum.  Since the civil construction of laboratories 
might contribute half or more to the cost of an experiment, finding existing
labs at those locations may change the optimization.


\par A list of possible sites for a new reactor experiment is
included in Table \ref{tab:sites}, along with a tabulation of
previous reactor experiment sites.
Two specific proposals have been made.  The KR2DET\cite{bib:kr2det}
experiment would use the Krasnoyarsk reactor in Russia, and two detectors
located at 115 and 1000 m.  That site has two attractive features:  1) There
is an entire city built at 600 MWE depth with possible locations for the
detector sites, and 2) The fuel cycle for that reactor is better understood
than for most reactors because the fuel is changed every few months rather than
years.  The other specific idea is the Kashiwazaki\cite{bib:kash} site in Japan, the location
of 7 reactors.
It is the most powerful reactor complex in the world, 24~Gw. The
reactors come in one cluster of 4 and one cluster of 3, so there would need
to be two near detectors and one far detector, possibly located in shafts
created by a large drill.  Other sites in the United States, France, Taiwan
and Brazil are being considered.  The site in Taiwan may be interesting because
of an existing road tunnel 2 km from the reactor.\cite{bib:taiwan}
\par 
A sensitivity of 0.02 in $\sin^2 2\quq$
can be achieved with as little as 250 ton-Gigawatt-years, while an
exposure of 8000 ton-Gigawatt-years may be required to achieve a
sensitivity of 0.01.\cite{bib:huber}
A two or more detector reactor experiment seems to be an attractive
option as part of the search for $\quq$.  It can probably find a non-zero
value for $\quq$ faster and less expensively than an off-axis experiment.
It does not face the degeneracies regarding CP parameters and the sign of
$\dmsq$, and hence cannot address those issues.  But a measurement of
$\quq$ by reactors followed by optimized off-axis experiments would together
measure neutrino parameters with much less uncertainty due to degeneracies
and correlations.

\begin{table}[h]
\begin{tabular}{|l|l|l|l|l|l|}\hline
Reactor & Location & L & Power & Overburden & Detector Mass \\ \hline \hline
Chooz & France & 1100 m & 8.5 Gw & 300 MWE & 5 ton \\ \hline
Bugey & France & 49/95 & 5.6 & 16 & 1/0.5 \\ \hline
Palo Verde & Arizona & 890 & 11.6 & 32 & 11.3 \\ \hline
KamLAND & Japan & $<180>$ & 200 (26) & 2700 & 1000 \\ \hline \hline
Krasnoyarsk & Russia & 115/1000 & 1 & 600 & 46 \\ \hline
Diablo Canon & California & $\sim$ 1 & 6.1 & 600 & \\ \hline
Wolf Creek & Kansas & $\sim$ 1 & 3.2 & & \\ \hline
Boulby & UK & 25 & 3.1 & 2860 & \\ \hline
Heilbronn & Germany & 19.5 & 6.4 & 480 & \\ \hline
Kashiwazaki & Japan & 1.7 & 24.3 & & 20 ton \\ \hline
Texono & Taiwan & 2.0 & 4.1 & & \\ \hline
Angra & Brazil & & 4.0  & & \\ \hline
IMB & Ohio & 10 & 1.2 & 1570  & \\ \hline\hline

\end{tabular}
\caption{Past Reactor Sites and Future Possibilities\label{tab:sites}}
\end{table}

\section{UNO}
\par One of the reasons for the tremendous progress in understanding
the neutrino has been the fact that several detectors were built 
underground to search
for nucleon decay.  They haven't found nucleon decay, but have made a number
of other discoveries.  Perhaps the lesson is that we should build yet another
detector to look for nucleon decay.  And maybe we will find that the
nucleon decays!
\par
\begin{figure}
\vspace*{13pt}
\begin{center}
\mbox{\epsfig{figure=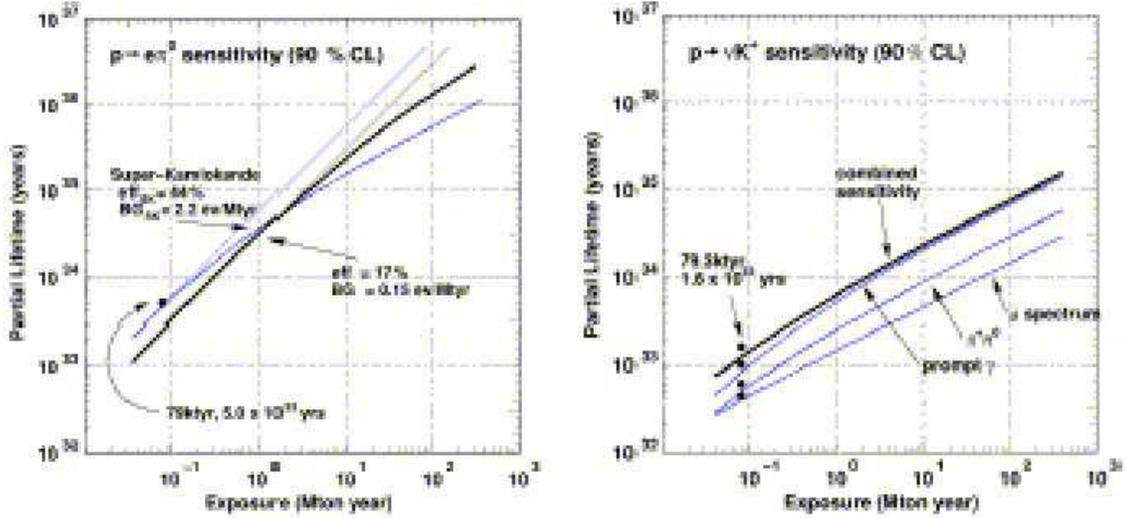,width=15.0cm}}
\caption{ Sensitivity versus exposure for UNO in the modes
$p \rightarrow \nu K^+$ and $p \rightarrow e \pi^0$.}
\end{center}
\label{fig:uno}
\end{figure}

The UNO detector\cite{bib:uno} is proposed as a next generation underground 
water Cerenkov detector that probes
nucleon decay beyond the sensitivities of the 
highly successful Super-Kamiokande (Super-K) detector utilizing a
well-tested technology. [1] The baseline 
conceptual design of the detector calls for a ``Multi-Cubical" design
with outer dimensions of 60x60x180 m3. The 
detector has three optically independent cubical
compartments with corresponding photo-cathode coverage 
of 10\%, 40\%, and 10\%, respectively. 
The total (fiducial) mass of the detector is 650 (440) kton,
which is about 13 (20) times larger than 
the Super-K detector. 
The discovery potential of the UNO detector is 
multi-fold. The probability of UNO discovering proton
decay via vector boson mediated $e^+\pi^0$
   mode is predicted to be quite high 
($\sim$50\%) in modern Grand
Unification Theories (GUTs). Water Cerenkov technology is the only 
realistic detector technology available today to allow a
search for this decay mode for proton lifetimes 
up to 10$^{35}$ years. If the current super-symmetric GUT
predictions are correct, UNO can discover proton 
decay via $\nu K^+$  mode.  The important design issue for any nucleon
decay detector is to keep backgrounds low while maintaining high efficiency.
UNO is able to take advantage of the understandings about these two modes that
have come from extensive analysis in the Super-Kamiokande experiment.
Modeling the backgrounds and efficiencies in an UNO sized detector, the 
sensitivity for these two modes is plotted versus exposure in 
Figure~\ref{fig:uno}.  While further advances are possible,  it is seen
that substantial increase in proton decay sensitivity would be achieved.
\par In addition to nucleon decay, UNO will be sensitive to
a large variety of other interesting topics.
UNO will be able to detect neutrinos 
from supernova explosions as far away as the Andromeda galaxy.
In case of a galactic supernova explosion, UNO 
will collect $\sim$100k neutrino events from which the
millisecond neutrino flux timing structure can be 
extracted. This could provide us with an observation of
black hole formation in real-time as well 
as a wealth of information to precisely determine the core
collapse mechanism.
Discovery of supernova relic neutrinos (SRN) is 
within the reach of UNO. SRN could very well be the
next astrophysical neutrinos to be discovered. The 
predicted values of the SRN flux by various theoretical
models are only up to six times smaller 
than the current Super-K limit. Some models have been already
excluded. With much larger fiducial mass and 
lower cosmogenic spallation background, UNO situated at a
depth 4000 MWE
can cover all of 
the predicted flux range. 
UNO is an ideal distant detector for a long-baseline 
neutrino oscillation experiment with neutrino beam
energies below about 10 GeV providing a synergy 
between the accelerator and the non-accelerator physics.
UNO provides other opportunities,
such as the ability to observe oscillatory behavior and   
appearance in the atmospheric neutrinos; precision measurement 
of temporal changes in the solar neutrino
fluxes; and searches for astrophysical point sources 
of neutrinos and dark matter in an energy range difficult
for larger, more coarse-grained undersea and under-ice detectors to cover.

\section{Summary}
There is a tremendous diversity of future neutrino projects and a number
of interesting measurements to be made, almost all of them involving
detectors located at facilities underground.   Even with the tremendous
progress we have already seen in the neutrino field, we are well on our
way to answering the questions that were posed in Section 1.
\section{Acknowledgments}
Thanks to Sandip Pakvasa and Chang Kee Jung for providing material for
this talk.   Thanks to Mike Shaevitz, Jerry Busenitz and Dave Reyna for
information on reactor neutrino projects.

\end{document}